\documentclass[aip,reprint]{revtex4-1}
\usepackage{url}
\usepackage{graphicx}
\usepackage{dcolumn}
\usepackage{bm}
\usepackage[T1]{fontenc}
\usepackage{etoolbox}
\usepackage{amsmath,amssymb}
\usepackage{booktabs}

\begin{document}


\title{Non-local physics-informed neural networks for forward and inverse solutions of granular flows} 



\author{Saghar Zolfaghari}
\affiliation{Department of Mechanical and Industrial Engineering, Northeastern University, Boston, MA, USA}

\author{Safa Jamali}
\email{s.jamali@northeastern.edu}
\affiliation{Department of Mechanical and Industrial Engineering, Northeastern University, Boston, MA, USA}


\date{\today}

\begin{abstract}
Dense granular flows exhibit nonlocal effects due to stress transmission in microplastic events, especially in quasi-static or slowly sheared regions. Hence, traditional local rheological models fail to capture spatial cooperativity effects that are prominent in many granular systems. The nonlocal granular fluidity (NGF) model addresses this limitation by introducing a diffusive-like partial differential equation for a fluidity field, governed by a key material-dependent parameter: the nonlocal amplitude \textit{A}. However, determining \textit{A} from experiments or simulations is known to be difficult and typically requires extensive calibration across multiple geometries. In this work, we present a data-driven platform based on Physics-Informed Neural Networks (PINNs) embedded with the NGF model, capable of solving granular flows in a forward or inverse manner. We show that once trained on transient flow fields, these non-local PINNs can readily infer the material parameters, as well as the pressure and stress fields. These data-driven frameworks allow for accurate recovery of small variations in the nonlocal amplitude, \textit{A}, which lead to sharp bifurcation-like transitions in the flow field. This approach demonstrates the feasibility of data-driven parameter inference in complex nonlocal models and opens up new possibilities for characterizing granular materials from sparse experimental observations.
\end{abstract}

\pacs{}

\maketitle 


Granular materials are ubiquitous in both everyday life and a wide range of industrial processes, including geotechnical engineering, energy, pharmaceuticals, and food processing \cite{henann2013predictive, Kamrin2014FrictionNonlocal, Jerolmack2019SoftMatterEarth, Coussot2005Rheometry, Forterre2008DenseGranular}. The mechanical response of dense granular packings spans from solid-like behavior at rest to fluid-like flow under applied stress, exhibiting rich and complex phenomenology. This complexity—further intensified by sensitivity to grain size, confinement, boundary conditions, and other rheological characteristics—poses a major challenge for conventional modeling approaches \cite{henann2013predictive, Kamrin2014FrictionNonlocal}. While traditional approaches such as rate-independent plasticity models from soil mechanics \cite{Schofield1968,Nedderman1992} and rate-dependent rheological models for rapid flows \cite{Jaeger1996,Halsey2002} have had success in their respective limits, they commonly fail to capture grain-size-dependent behavior and transitional regimes where solid- and fluid-like responses coexist \cite{Henann2014Thermo}.

To bridge the gap between solid-like and fluid-like descriptions, Jop and coworkers combined rate-dependent flow models with yield criteria, leading to a unified rheological framework based on the dimensionless relation $\mu=\mu(I)$, where the shear stress–to–pressure ratio depends on the inertial number $I$~\cite{Jop2006ConstitutiveLaw, DaCruz2005Rheophysics, PouliquenForterre2002Friction}. Supported by dimensional analysis and experimental validation, this approach successfully describes granular behavior in simple, uniform flows such as planar shear~\cite{henann2013predictive, Andreotti2013GranularMedia}. However, in slowly flowing, quasi-static regimes—typically characterized by $I<10^{-3}$ and non-uniform geometries—the one-to-one relationship between $\mu$ and $I$ breaks down, and local rheological models fail to capture the observed behavior~\cite{Kamrin2014FrictionNonlocal, henann2013predictive, Bouzid2013NonlocalYield, Bouzid2015FluidityReview}. In such cases, local rheological models, which describe granular materials solely in terms of local stress and strain-rate conditions, while effective for simple, uniform flows, become inadequate for spatially heterogeneous deformation, such as creeping zones or regions near boundaries. This breakdown reflects the inherently nonlocal nature of granular flow in these regimes, where particle interactions and momentum transfer extend beyond the immediate neighborhood. Accurate modeling therefore requires nonlocal formulations that introduce an internal length scale—often proportional to the grain diameter—to capture spatial cooperativity and geometric sensitivity~\cite{kamrin2019nonlocality, Kamrin2024Review, Pouliquen2009Nonlocal, Tang2018NonlocalAnnular}.

Various modeling strategies have been proposed to capture nonlocal effects in granular materials, including Cosserat continua~\cite{Mohan2002}, gradient plasticity models~\cite{Hashiguchi2007}, convolution-based integral formulations~\cite{Pouliquen2009}, kinetic theory approaches~\cite{Jenkins2010, Savage1998}, and nonlocal order parameter models~\cite{Bouzid2013, Aranson2002, Kamrin2007}. These frameworks aim to introduce an internal length scale that reflects the spatial cooperativity of granular interactions~\cite{kamrin2019nonlocality}. Among these models, one prominent approach is the Nonlocal Granular Fluidity (NGF) model, which introduces a granular fluidity field, denoted by \( g \), a scalar state variable that quantifies the material’s capacity to flow under small applied stresses. The fluidity is typically defined as the ratio of shear rate to shear stress, \( g = \dot{\gamma}/\mu \), effectively acting as a pressure-weighted inverse viscosity~\cite{kamrin2019nonlocality}, and evolves according to a diffusion-like partial differential equation~\cite{Kamrin2012, henann2013predictive}:

\begin{equation}
    t_0 \,\dot{g} = A^2 d^2 \nabla^2 g - \Big[(\mu_s - \mu)g + b \sqrt{\frac{\rho_s d^2}{P}} \,\mu g^2\Big].
    \label{eq:ngf}
\end{equation}

Here, \( t_0 \) is a characteristic time scale, \( A \) is the dimensionless nonlocal amplitude, \( d \) is the particle diameter, \( b \) characterizes the rate-dependent response, \( \mu_s \) is the static yield threshold, \( \rho_s \) is the solid density, and \( P \) is the pressure. The governing equation balances the temporal evolution of the fluidity field with spatial diffusion and nonlinear source terms. The Laplacian diffusion term, scaled by \( A \) and \( d \), introduces a grain-size-dependent cooperativity length that enables spatial propagation of fluidity, allowing it to extend into nearby sub-yield regions and giving rise to finite-width shear bands—phenomena inaccessible to purely local rheological models~\cite{kamrin2012nonlocal}. The source terms, involving the stress ratio \( \mu \), pressure \( P \), and friction coefficient \( \mu_s \), regulate transitions between flowing and static states, while the quadratic \( g^2 \) term captures rate sensitivity and nonlinear amplification near actively flowing regions~\cite{Henann2014, Kamrin2014FrictionNonlocal}. This formulation has been successfully applied across planar shear, annular Couette flow, and inclined-plane geometries, and is supported by extensive experimental and numerical evidence~\cite{Tang2018NonlocalAnnular, Kamrin2024Review, henann2013predictive, kamrin2012nonlocal}, with further validation provided by correlations between fluidity and kinematic observables such as velocity fluctuations and packing density~\cite{zhang2017microscopic}. Consequently, these models capture rich dynamical behavior, including rate-dependent strengthening, size-dependent “smaller-is-stronger” effects arising from reduced cooperativity, and secondary rheology, wherein shear-induced fluidization in one region enables flow or creep in distant, nominally static zones~\cite{kamrin2019nonlocality, Henann2014, Poon2023MicroscopicGranularFluidity, Haeri2022MPM_NGF, Clarke2024NGF_NMR}. Through its spatial coupling mechanism, NGF captures key nonlocal phenomena, including flow in sub-yield regions and finite-width shear band formation, which remain inaccessible to conventional local rheological models, making the fluidity field a physically grounded variable for representing nonlocal behavior in dense, complex materials~\cite{kamrin2019nonlocality, kamrin2012nonlocal}.

Regardless of the specific nonlocal formulation, models for dense granular flow hinge on defining an amplitude that controls the decay of nonlocal effects. This nonlocal amplitude quantifies the extent of spatial cooperativity, which governs the diffusion of the fluidity field, $g$, and determines how far the influence of a local shear zone propagates through a granular medium~\cite{henann2013predictive, Kamrin2012}. This length scale is essential for capturing flow in sub-yield regions, finite-width shear bands, and rate-independent creep (i.e., secondary rheology), all of which lie beyond the scope of local rheological models~\cite{Kamrin2014FrictionNonlocal, Henann2014}. Despite its central role, determining this amplitude—denoted $A$ in the NGF framework—remains a key limitation. It cannot be measured directly, even in numerical simulations, because it appears as the coefficient of a second-order spatial derivative in a partial differential equation governing an implicit state variable, namely the granular fluidity, which itself is not directly observable~\cite{Kamrin2014FrictionNonlocal, zhang2017microscopic}. Consequently, existing approaches rely on multi-step calibration procedures involving discrete element method (DEM) simulations in specific geometries (e.g., annular Couette flow), followed by fitting NGF predictions to measured flow profiles~\cite{henann2013predictive, Kamrin2012, Kamrin2014FrictionNonlocal}. Such procedures are computationally intensive, geometry dependent, and often lack transferability to new flow configurations. In contrast to continuum parameters such as the static friction coefficient $\mu_s$ or rate sensitivity $b$, which can often be extracted from uniform shear tests, $A$ encodes inherently nonlocal, system-scale behavior and therefore depends on global flow structure, boundary conditions, and geometry. Although recent efforts have sought to relate $A$ to grain-scale properties such as particle friction or velocity fluctuations~\cite{Kamrin2014FrictionNonlocal, zhang2017microscopic}, a practical and general method for estimating $A$ without extensive calibration remains an open challenge, limiting the predictive applicability of the NGF model in untested geometries. From a computational standpoint, solving the resulting coupled system of governing equations using conventional numerical methods, such as finite element or finite volume approaches, is often expensive and difficult to generalize across complex geometries and boundary conditions~\cite{Mahmoudabadbozchelou2021, Mahmoudabadbozchelou2022nnPINNs}. Moreover, parameter calibration—most notably of the nonlocal amplitude $A$—typically requires DEM simulations or carefully controlled experiments combined with geometry-specific fitting procedures~\cite{henann2013predictive, Kamrin2014FrictionNonlocal}.

A Physics-Informed Neural Network (PINN) is a class of machine-learning models that incorporates governing physical laws directly into the training process by embedding the residuals of the underlying continuum equations into the loss function~\cite{Mahmoudabadbozchelou2021, Mahmoudabadbozchelou2022nnPINNs}. Unlike purely data-driven approaches, PINNs enforce physical consistency while simultaneously learning field variables and unknown model parameters, without requiring labeled data for internal quantities such as stress or fluidity. Recent developments in physics-informed learning include fractional-order formulations for systems with memory effects and anomalous transport~\cite{Dabiri2023RhINN, Dabiri2025fPINNReview}, neural-operator-based methods for learning solution operators across varying conditions~\cite{saberi2025rheoformer}, and particle-scale machine-learning approaches for inferring frictional contact networks in dense suspensions~\cite{Aminimajd2025FCN, Aminimajd2025FCNarXiv}. Within this broader landscape, PINNs are particularly well suited for nonlocal continuum models, where governing equations are known but key parameters remain inaccessible to direct measurement, allowing quantities such as the nonlocal amplitude A to be treated as learnable global parameters constrained by physics.

In this work, we develop a two-stage PINN framework to infer the nonlocal amplitude \( A \) in dense granular flows governed by the NGF model using only velocity observations. In the first stage, a direct PINN is trained with a prescribed value of \( A \) to solve the full coupled system of governing equations, yielding velocity and stress fields that are fully consistent with NGF predictions and validated against an optimized basis-expansion numerical solver. In the second stage, an inverse PINN is trained using velocity data alone, with \( A \) embedded directly in the physics-constrained loss function as a trainable global parameter. By enforcing all governing equations without requiring labeled stress or fluidity fields, this approach enables accurate recovery of \( A \) with sub-percent error across a range of test cases. 

\begin{figure*}[htbp]
    \centering
    \includegraphics[width=0.95\textwidth]{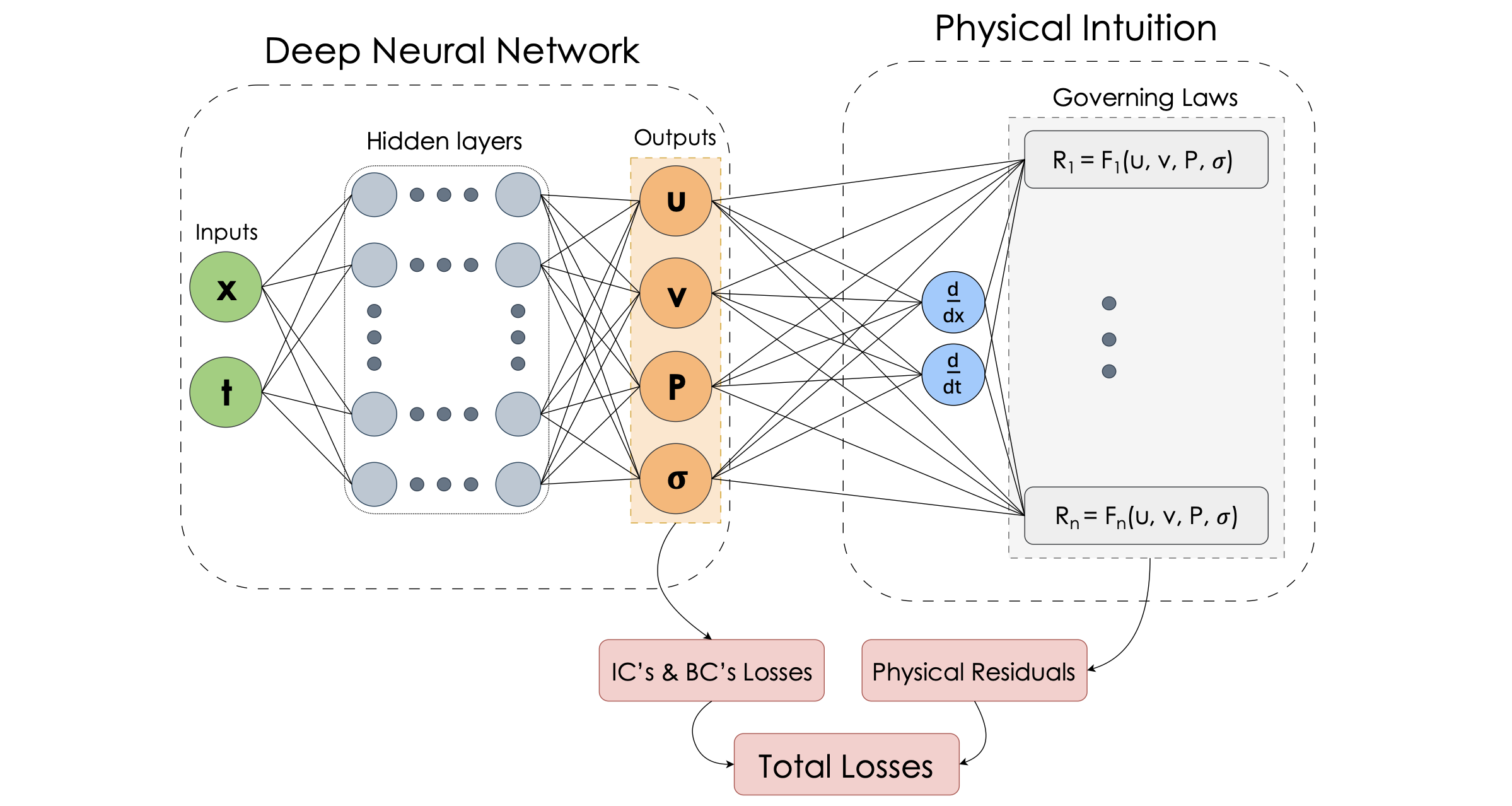}
    \caption{Schematic of the Physics-Informed Neural Network (PINN) framework developed in this work for structured fluid modeling. The architecture consists of a deep neural network coupled with embedded physical laws. Spatial and temporal coordinates \( (x, y, t) \) are provided as inputs, and the network predicts physical fields including velocity \( (u, v) \), pressure \( P \), and stress components \( (\sigma_{xy}, \sigma_{xx}, \sigma_{yy}) \). Automatic differentiation is used to compute the required spatial and temporal derivatives, which are substituted into the governing equations \( F_i(\sigma, P, u, v) \) representing momentum conservation, incompressible continuity, and nonlocal fluidity evolution. The resulting residuals, together with initial and boundary condition constraints, form the total loss function guiding training and enforcing physical consistency.}
    \label{fig:myimage}
\end{figure*}

The proposed PINN architecture provides a unified, mesh-free framework for solving the full system of governing equations in dense granular flows. Spatial and temporal coordinates \( (x, y, t) \) are used as inputs, from which the network predicts the complete set of physical fields, including velocity components \( u \) and \( v \), pressure \( P \), and stress components \( \sigma_{xx} \), \( \sigma_{yy} \), and \( \sigma_{xy} \). All spatial and temporal derivatives required by the governing equations are evaluated using automatic differentiation, enabling accurate, mesh-free computation of PDE residuals without explicit numerical discretization. By embedding the nonlocal fluidity equation together with the incompressible continuity and momentum balance equations into the training process, the PINN naturally captures spatial cooperativity and resolves key nonlocal phenomena, including shear localization, sub-yield creep, and finite-width deformation bands, which are inaccessible to conventional local rheological models.

Rather than relying on labeled training data, physical consistency is enforced through a physics-informed loss function constructed directly from the governing equations. Training proceeds by minimizing a composite loss composed of the residuals of each governing equation, together with soft penalty terms enforcing the prescribed initial and boundary conditions. Specifically, the total physics-informed loss is defined in terms of a set of residual functions \( f_i(x,y,t) \), each corresponding to a governing equation, and augmented by initial-condition (IC) and boundary-condition (BC) losses:

\begin{equation}
\text{MSE}_R = \sum_{j=1}^{N_{\text{eqs}}} \frac{1}{N_{R_j}} \sum_{i=1}^{N_{R_j}} 
\left| \text{Residual}_{\text{eq}_j}(t_i) \right|^2
\end{equation}

\begin{equation}
\begin{aligned}
\mathrm{MSE}_{\mathrm{IC}}
&=
\sum_{j=1}^{N_{\mathrm{outputs}}}
\frac{1}{N_{\mathrm{IC},j}}
\sum_{i=1}^{N_{\mathrm{IC},j}}
\\
&\quad
\left|
\mathrm{Predicted}\!\left(u_{j,\mathrm{IC},i}\right)
-
\mathrm{Actual}\!\left(u_{j,\mathrm{IC},i}\right)
\right|^2
\end{aligned}
\end{equation}

\begin{equation}
\begin{aligned}
\mathrm{MSE}_{\mathrm{BC}}
&=
\sum_{j=1}^{N_{\mathrm{outputs}}}
\frac{1}{N_{\mathrm{BC},j}}
\sum_{i=1}^{N_{\mathrm{BC},j}}
\\
&\quad
\left|
\mathrm{Predicted}\!\left(u_{j,\mathrm{BC},i}\right)
-
\mathrm{Actual}\!\left(u_{j,\mathrm{BC},i}\right)
\right|^2
\end{aligned}
\end{equation}

and the total loss is given by
\begin{equation}
\mathrm{MSE} = \mathrm{MSE}_R + \omega_2 \mathrm{MSE}_{\mathrm{IC}} + \omega_3 \mathrm{MSE}_{\mathrm{BC}}.
\end{equation}

For both the direct and inverse PINNs, a fully connected feedforward neural network architecture is employed, consisting of six hidden layers with 60 neurons per layer and hyperbolic tangent (\(\tanh\)) activation functions. The network outputs include the velocity components \((u, v)\), pressure \(P\), and stress components \((\sigma_{xx}, \sigma_{yy}, \sigma_{xy})\). In the inverse PINN formulation, the nonlocal amplitude \(A\) is introduced as an additional trainable scalar parameter, allowing its value to be inferred directly from velocity observations. All network weights are initialized using the Glorot (Xavier) initialization scheme to promote stable gradient propagation during training. Following the weighting strategy introduced in the nn-PINN framework~\cite{Mahmoudabadbozchelou2022nnPINNs}, the initial-condition (IC) and boundary-condition (BC) losses are scaled using \(\omega_2 = 50\) and \(\omega_3 = 20\), respectively. This choice ensures strong enforcement of physically meaningful initial and boundary conditions while preserving the network’s ability to accurately resolve the internal dynamics governed by the PDEs.

The direct PINN is trained using a two-stage optimization strategy, consisting of an initial phase with the Adam optimizer at a learning rate of \(10^{-3}\) for 10{,}000 iterations, followed by fine-tuning with the L-BFGS-B algorithm. This combination improves convergence and enhances the accuracy of the predicted physical fields. In contrast, the inverse PINN is trained exclusively using the Adam optimizer for up to 40{,}000 iterations, during which both the network weights and the nonlocal amplitude \(A\) are optimized simultaneously. This choice avoids potential instabilities associated with quasi-Newton optimization when coupling network parameters with a global scalar variable.

Throughout training, the loss function enforces incompressible continuity, momentum balance, and the transient nonlocal fluidity equation, together with the prescribed initial and boundary conditions. These conditions are specified according to the two canonical flow configurations considered in this study: two-dimensional planar shear flow and two-dimensional pressure-driven flow. In the shear-driven configuration, the flow is confined between two parallel horizontal plates. A no-slip condition is imposed at the lower boundary (\(y = 0\)), enforcing \(u = 0\) and \(v = 0\), while a unit tangential velocity (\(u = 1\)) is prescribed at the upper boundary (\(y = 1\)) to drive shear. The system is initialized from rest, with zero velocity throughout the domain at \(t = 0\). In the pressure-driven configuration, a constant horizontal pressure gradient drives the flow, with no-slip conditions imposed at both the upper and lower boundaries and zero initial velocity. In this case, no velocity is prescribed on the boundaries, and the flow develops solely in response to the imposed pressure gradient.

To accurately represent gravity-driven granular flows, body forces due to gravity are explicitly included in the \( y \)-momentum equation. This contribution is essential for capturing nonlocal effects such as sub-yield creep and shear-band formation, which have been shown to emerge only when gravitational loading is accounted for~\cite{zhang2017microscopic, kamrin2012nonlocal}. Although shear is applied in the horizontal direction, gravity acting in the vertical direction plays a central role in stress redistribution and flow behavior in dense granular systems. Accordingly, the governing flow equations consist of the incompressible continuity equation and the conservation of linear momentum:

\begin{equation}
\left\{
\begin{aligned}
\nabla \cdot \mathbf{v} &= 0, \\
\frac{D\mathbf{v}}{Dt} &= -\frac{1}{\rho} \nabla P - \frac{1}{\rho} \nabla \cdot \boldsymbol{\sigma} + \mathbf{g_y},
\end{aligned}
\right.
\end{equation}

Here, \( \mathbf{v} = (u, v) \) is the velocity field, \( P \) is the pressure, \( \boldsymbol{\sigma} \) is the Cauchy stress tensor, \( \rho \) is the bulk density, and \( \mathbf{g_y}{} \) denotes the gravitational acceleration vector. The stress tensor is governed by a nonlocal constitutive relation involving a diffusive fluidity field rather than a local strain-rate law. In two-dimensional Cartesian coordinates, the governing equations take the form
\begin{equation}
\left\{
\begin{aligned}
\frac{\partial u}{\partial x} + \frac{\partial v}{\partial y} &= 0, \\
\frac{\partial u}{\partial t} + u \frac{\partial u}{\partial x} + v \frac{\partial u}{\partial y} &=
-\frac{1}{\rho} \frac{\partial P}{\partial x}
-\frac{1}{\rho} \left( \frac{\partial \sigma_{xx}}{\partial x} + \frac{\partial \sigma_{xy}}{\partial y} \right), \\
\frac{\partial v}{\partial t} + u \frac{\partial v}{\partial x} + v \frac{\partial v}{\partial y} &=
-\frac{1}{\rho} \frac{\partial P}{\partial y}
-\frac{1}{\rho} \left( \frac{\partial \sigma_{yx}}{\partial x} + \frac{\partial \sigma_{yy}}{\partial y} \right)
+ g_y,
\end{aligned}
\right.
\end{equation}


In the first step, a direct PINN is trained using the full set of governing equations with prescribed rheological parameters, and its predictions are validated against numerical solutions obtained using a spectral basis-expansion method for two planar benchmark configurations: (i) shear-driven flow and (ii) pressure-driven flow. We then employ an inverse PINN to infer the nonlocal amplitude \(A\) using velocity observations alone. In addition, we examine the influence of the key rheological parameters \(A\), \(\mu_s\), and \(b\) on the resulting flow fields, highlighting their roles in controlling shear localization and spatial cooperativity.

\begin{figure*}[htbp]
    \centering

    \includegraphics[trim=0 0 17 0, clip, width=0.9\textwidth]{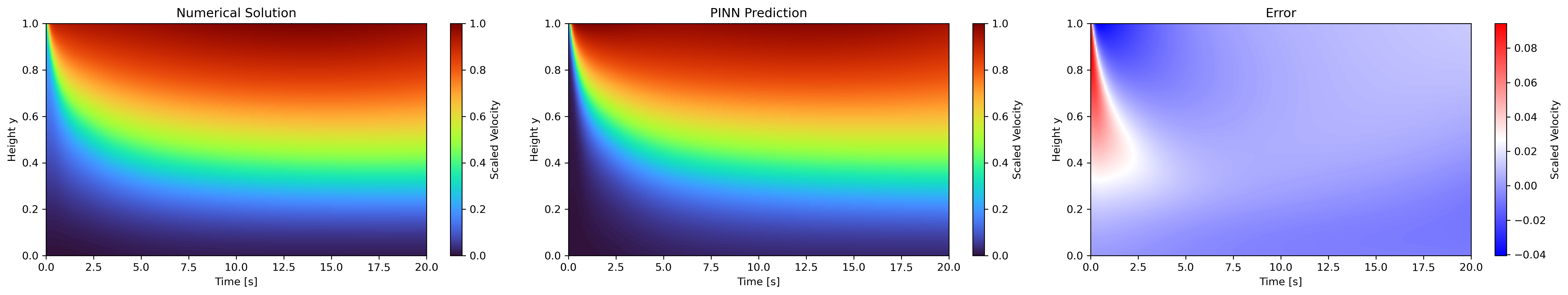}
    \vspace{1em}
    
    \includegraphics[trim=0 0 17 0, clip, width=0.9\textwidth]{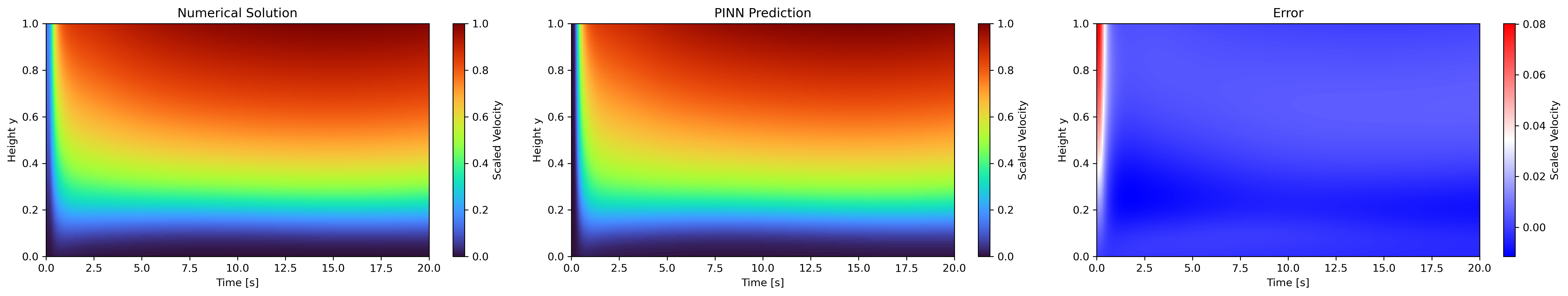}

    \caption{Comparison of velocity profiles and absolute errors between numerical solutions and Direct PINN predictions for dense granular flow governed by NGF in a planar shear configuration. The top and bottom panels correspond to nonlocal amplitude values of \( A = 0.97 \) and \( A = 1.05 \), respectively. Simulations are performed with solid density \( \rho_s = 2450\, \mathrm{kg/m^3} \), particle diameter \( d = 0.0008\, \mathrm{m} \), no-slip boundary conditions, and zero initial velocity. The first column shows the reference velocity field obtained from a spectral numerical solver, the second column presents PINN prediction, and the third column displays the point-wise absolute error.}
    \label{fig:shear_direct}
\end{figure*}

Figure~\ref{fig:shear_direct} shows the predicted velocity fields together with the corresponding error maps for two representative values of the nonlocal amplitude, \(A = 0.97\) and \(A = 1.05\), in the shear-driven configuration. In both cases, the direct PINN accurately captures the formation, localization, and decay of the shear band near the driven boundary, with uniformly low errors across the spatiotemporal domain. Even small variations in \(A\) lead to pronounced changes in the velocity profiles, yield surface, and transient response, underscoring the strong sensitivity of the flow dynamics to the nonlocal amplitude.

\begin{figure*}[htbp]
    \centering

    \includegraphics[trim=0 0 17 0, clip, width=0.9\textwidth]{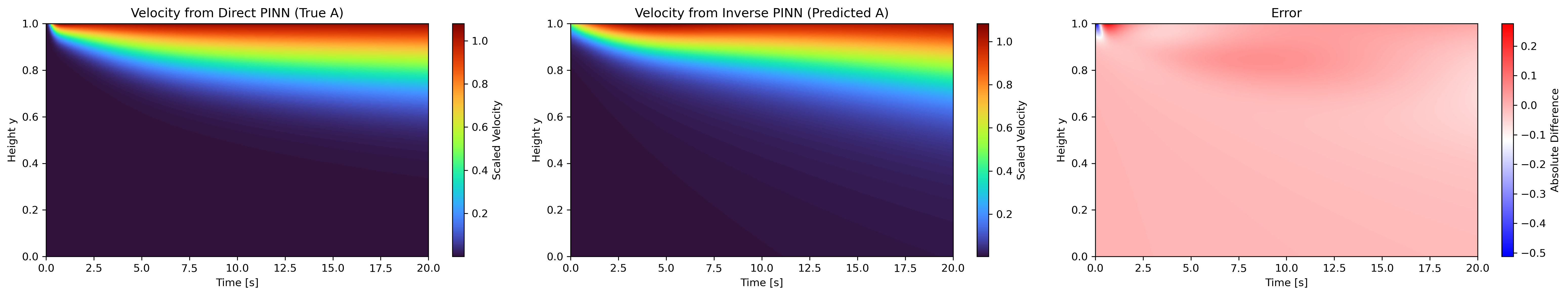}
    \vspace{1em}

    \includegraphics[trim=0 0 17 0, clip, width=0.9\textwidth]{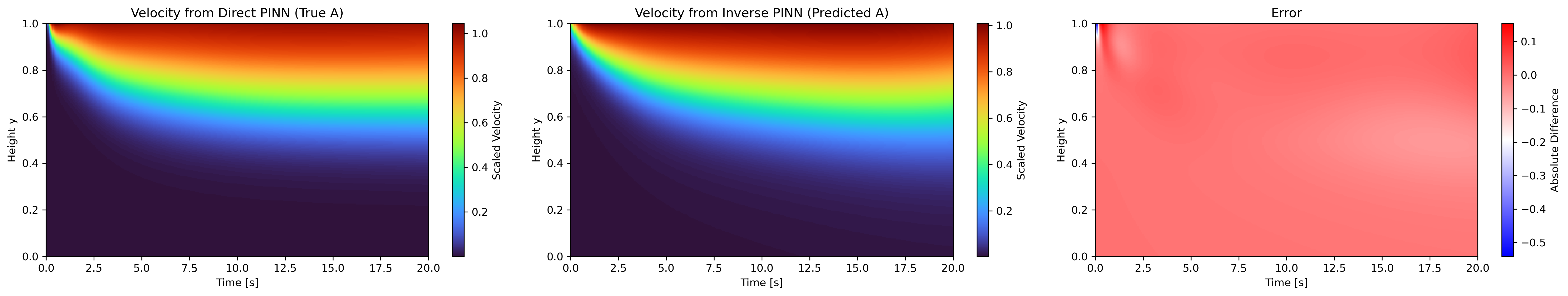}
    \vspace{1em}

    \includegraphics[trim=0 0 17 0, clip, width=0.9\textwidth]{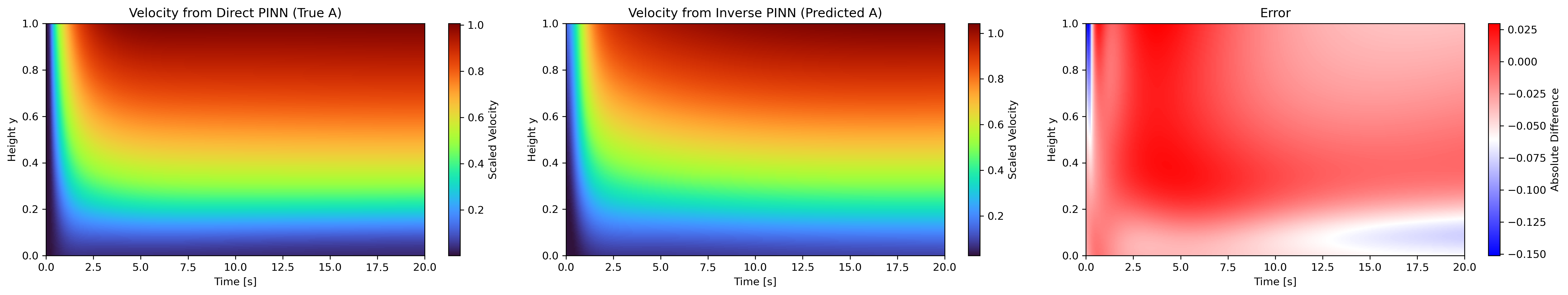}

    \caption{Comparison of velocity profiles and corresponding absolute errors between forward and inverse PINN predictions for three test cases in the shear-driven configuration: \( A = 0.84 \) (top), \( A = 0.85 \) (middle), and \( A = 1.03 \) (bottom). The model is trained using only velocity observations and the governing equations, with no labeled stress or fluidity data. All simulations assume a two-dimensional domain with solid density \( \rho = 2450~\mathrm{kg/m^3} \) and particle size \( d = 0.0008~\mathrm{m} \). The first column shows the reference velocity field, the second column presents the Inverse PINN prediction, and the third column shows the point-wise absolute error.}
    \label{fig:shear_inverse}
\end{figure*}

To assess the inverse problem, we consider the same shear-driven configuration while treating \(A\) as an unknown global parameter. The inverse PINN is trained using velocity data only, with no labels provided for stress or fluidity fields, while the remaining rheological parameters \(\mu_s\) and \(b\) are fixed. Figure~\ref{fig:shear_inverse} presents representative results for three cases (\(A = 0.84\), \(0.85\), and \(1.03\)), showing accurate reconstruction of both the shear-band structure and the full flow evolution, with small residuals throughout the domain. Table~\ref{tab:A_comparison} summarizes the inferred values of \(A\) across multiple parameter sets. In all cases, the inverse PINN recovers the correct nonlocal amplitude with sub-percent relative error, demonstrating robust and data-efficient parameter inference.

\begin{table}[h]
\centering
\caption{Inverse PINN estimation of the nonlocal amplitude \( A \) across various test cases. Each row reports the true and predicted values of \( A \), along with the corresponding static friction coefficient \( \mu_s \) and rate sensitivity parameter \( b \) used in the simulation.}
\label{tab:A_comparison}
\begin{tabular}{cccc}
\toprule
True \(A\) & Predicted \(A\) & \(\mu_s\) & \(b\) \\
\midrule
0.83 & 0.8307 & 0.205 & 1.35 \\
0.84 & 0.8410 & 0.290 & 1.50 \\
0.85 & 0.8498 & 0.245 & 1.35 \\
1.03 & 1.0314 & 0.145 & 1.20 \\
1.05 & 1.0492 & 0.106 & 1.25 \\
\bottomrule
\end{tabular}
\end{table}

\begin{figure*}[htbp]
    \centering

    \includegraphics[trim=0 0 17 0, clip, width=0.9\textwidth]{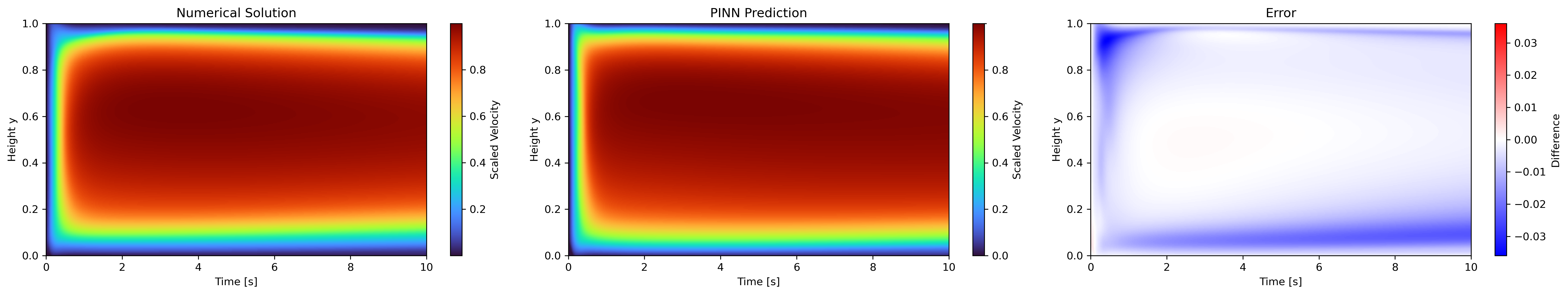}

    \caption{Comparison of velocity profiles and absolute errors between numerical solutions and Direct PINN predictions for dense granular flow governed by the Nonlocal Granular Fluidity (NGF) model in a pressure-driven configuration. The simulation domain is two-dimensional with no-slip frictional boundaries at the top and bottom, and flow is driven by a horizontal pressure gradient. The material parameters are density \( \rho_s = 2450\, \mathrm{kg/m^3} \) and particle diameter \( d = 0.0008\, \mathrm{m} \), with an initial condition of zero velocity throughout the domain. The first column shows the reference velocity field obtained from a numerical spectral solver, the second column presents the Direct PINN prediction trained with known rheological parameters, and the third column displays the pointwise absolute error. The nonlocal amplitude used in these simulations is \( A = 0.48 \).}
    \label{fig:pressure_direct}
\end{figure*}

In the pressure-driven configuration, we consider a two-dimensional planar domain in which a horizontal pressure gradient drives the flow, generating a velocity profile across the material. This configuration provides a stringent test of the NGF model under boundary conditions and flow geometries representative of confined granular systems, such as channels or silos. The system is initialized from a quiescent state with zero velocity throughout the domain. For this case, we employ a direct PINN trained with known rheological parameters, specifically \(A = 0.48\), \(\mu_s = 0.382\), and \(b = 0.938\), using the same NGF formulation as in the shear-driven configuration. The PINN learns the spatiotemporal velocity, stress fields, and the internal pressure distribution, directly from the governing equations without requiring labeled data. Figure~\ref{fig:pressure_direct} presents the predicted velocity profiles and the corresponding point-wise errors relative to reference solutions obtained using a spectral numerical solver. As shown in this figure, the PINN accurately resolves the pressure-driven flow, capturing both the velocity maximum near the channel center and the decay toward the boundaries due to frictional resistance, with uniformly small errors across the domain.

\begin{figure*}[htbp]
    \centering

    \includegraphics[trim=0 0 17 0, clip, width=0.9\textwidth]{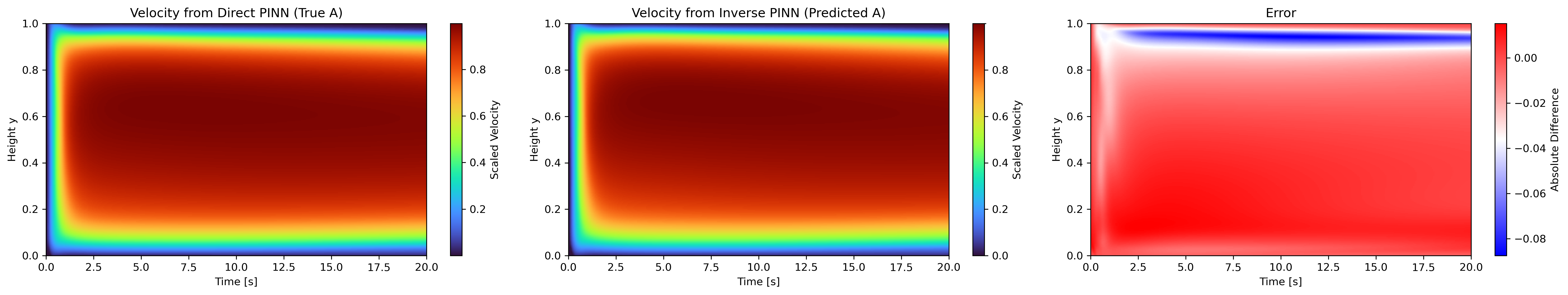}

    \caption{Comparison of velocity profiles and absolute errors from direct and inverse PINN models in a pressure-driven NGF flows. The simulation domain is two-dimensional with frictional no-slip boundaries and a horizontal pressure gradient driving the flow, using a density \( \rho = 2450~\mathrm{kg/m^3} \) and particle diameter \( d = 0.0008~\mathrm{m} \). The first panel shows the reference velocity field from forward PINN (trained with known rheological parameters), the second panel presents inverse PINN prediction (using only velocity observations and treating \( A \) as an unknown), and the third panel displays the point-wise absolute error. The inferred value \( A = 0.476 \) closely matches the true value \( A = 0.48 \), confirming the effectiveness of the inverse framework under dynamic pressure-driven conditions.}
    \label{fig:pressure_inverse}
\end{figure*}

We next perform inverse modeling to assess whether the nonlocal amplitude \(A\) can be inferred from velocity data alone in the pressure-driven configuration. For the test case considered, the inverse PINN predicts \(A = 0.476\), in close agreement with the reference value used in the direct simulation (\(A = 0.48\)). The resulting velocity profiles remain in excellent agreement with the NGF predictions, confirming the accuracy and stability of the inverse formulation. Figure~\ref{fig:pressure_inverse} shows the predicted velocity field together with the corresponding absolute error relative to the numerical reference solution. Together, these results demonstrate that the proposed framework generalizes robustly across distinct flow regimes and enables reliable recovery of latent nonlocal parameters under pressure-driven boundary conditions.

The nonlocal amplitude \(A\) controls the strength of spatial diffusion in the granular fluidity equation and thus sets the degree of cooperativity in the NGF model. By scaling the diffusion term, \(A\) determines how far local flow activity propagates into neighboring regions. Larger \(A\) enhances nonlocal coupling, leading to deeper fluidity penetration into sub-yield zones and wider shear bands with smoother velocity gradients, whereas smaller \(A\) confines deformation near the yield surface. This trend is evident in Fig.~\ref{fig:shear_direct}: for \(A = 1.05\), the shear band extends broadly into the bulk, while for \(A = 0.97\), deformation remains sharply localized near the upper boundary, consistent with NGF theory.

The dimensionless parameter \(b\) governs rate sensitivity through the nonlinearity of the fluidity source term above the yield threshold \(\mu_s\). Larger \(b\) produces sharper velocity gradients and stronger localization by amplifying fluidity abruptly in actively flowing regions, whereas smaller \(b\) allows smoother transitions and broader shear zones. This behavior is observed in the inverse PINN results (Fig.~\ref{fig:shear_inverse}): the case with \(b = 1.50\), \(A = 0.84\), and \(\mu_s = 0.290\) exhibits a tightly confined shear band, while \(b = 1.20\), \(A = 1.03\), and \(\mu_s = 0.145\) yields a broader, more gradual shear profile.

The static friction coefficient \(\mu_s\) sets the local yield threshold and controls flow initiation. Larger \(\mu_s\) suppresses fluidity in sub-yield regions, resulting in narrow shear layers, whereas smaller \(\mu_s\) lowers the yielding threshold and promotes wider regions of active shear. Accordingly, the velocity field remains steep and localized for \(\mu_s = 0.290\) and \(A = 0.84\), while for \(\mu_s = 0.106\) with \(A = 1.05\), the shear zone broadens significantly, indicating easier flow onset and deeper fluidity penetration.

Taken together, these results demonstrate that the NGF parameters \(A\), \(b\), and \(\mu_s\) exert distinct and physically interpretable controls on dense granular flow behavior. Across both shear-driven and pressure-driven configurations, the PINN solutions consistently reproduce the expected trends: larger values of \(A\) and smaller values of \(b\) or \(\mu_s\) promote enhanced nonlocal cooperativity and broader, more distributed shear zones, whereas the opposite parameter combinations lead to sharply localized deformation. Importantly, these behaviors are captured not only by the direct PINN simulations but are also accurately recovered through inverse PINN inference using velocity data alone. The close agreement between forward PINN predictions, inverse parameter recovery, and established NGF theory underscores the robustness of the proposed PINN framework as a unified tool for both predictive modeling and data-efficient identification of nonlocal rheological parameters in dense granular flows.


In this study, we developed a physics-informed neural network (PINN) framework for modeling dense granular flows governed by the Nonlocal Granular Fluidity (NGF) model. By embedding the full system of governing equations—including incompressible continuity, momentum conservation, and the transient nonlocal fluidity equation—directly into the network, we obtained a unified, mesh-free approach that learns physically consistent solutions from sparse velocity data without requiring labeled internal stress or fluidity fields. Our two-stage strategy combines a direct PINN, trained with a known nonlocal amplitude $A$, and an inverse PINN that infers $A$ solely from velocity observations. The direct model was validated against a numerical spectral solver, showing excellent agreement across the spatiotemporal domain. As reported in Table~\ref{tab:A_comparison}, the inferred values of $A$ exhibit sub-percent relative errors across a wide range of test cases with varying $\mu_s$ and $b$, typically below $0.2\%$, demonstrating robust and efficient recovery of latent rheological parameters in a physics-constrained setting. The benchmarking results further show that the velocity field is highly sensitive to small variations in $A$, highlighting its central role in nonlocal flow behavior.

Together, these results underscore the significance of integrating PINNs with nonlocal granular fluidity models for parameter discovery in complex rheologies. To the best of our knowledge, the proposed nonlocal PINN framework provides the first practical route for connecting experimentally observable velocity fields to detailed nonlocal continuum models and their parameters. Unlike traditional calibration approaches based on geometry-specific DEM simulations, the method generalizes across flow configurations and avoids the need for fully resolved flow-field data. Incorporation of gravity-induced body forces enables realistic modeling of flows and reveals how gravitational gradients enhance spatial cooperativity through the nonlocal fluidity mechanism. Enforcing the nonlocal PDE as a hard constraint ensures that the learned solutions capture both local constitutive behavior and long-range cooperative effects, allowing accurate representation of shear banding and sub-yield creep.

Overall, this work establishes a robust, interpretable, and scalable alternative to conventional solvers and fitting-based calibration workflows. By enabling accurate recovery of hidden internal fields and material parameters from minimal data, the proposed PINN framework offers a powerful tool for studying nonlocal rheology in granular systems. The approach can be readily extended to multi-parameter inference, boundary-condition calibration, three-dimensional domains, irregular geometries, and experimental datasets.




\bibliographystyle{aipnum4-1}
\bibliography{References}

\end{document}